\documentstyle[preprint,prb,aps]{revtex}
\begin{document}
\draft
\title{Occupation numbers in the density functional
calculations.}
\author{M.M.Valiev and G.W. Fernando}
\address{Physics Department, University of Connecticut,
Storrs, Connecticut, 06269, USA}
\date{April 26, 1995}
\maketitle
\begin{abstract}
It is the intention of this paper to rigorously clarify the role
of the occupation numbers in the current practical applications
of the density functional formalism.
In these calculations one has to decide how to distribute
a given, fixed number of electrons over a set
of single-particle orbitals.
The conventional choice
is to have single-particle orbitals
below the Fermi level completely occupied (with possible fractional
occupations at the Fermi level) and the orbitals
above the Fermi level empty.
Although there is a certain confusion in the literature why this
choice is superior to any others, the general belief is that it
can be justified by treating the occupation numbers
as variational parameters and then applying
Janak's theorem or similar reasoning.
 We demonstrate that there is a serious flaw
in those arguments, mainly the kinetic
energy functional and therefore the exchange-correlation functional
are not differentiable with respect to the density for arbitrary
occupation numbers.
It is rigorously shown that
in the present context of the density
functional calculations there is no freedom to vary the occupation
numbers. The occupation numbers cannot be considered
as variational parameters.
\end{abstract}
\pacs{71.10}
\section{Introduction.}
Predicting properties of real materials from first principles
has been an important goal of solid state physics from the
early days, going back to Thomas, Fermi and others. During the
past few decades, density functional theory \cite{hk,ks,lp}
 (DFT) within the local
density approximation (LDA) has been quite successful in this
regard. This method has no adjustable parameters and
the only significant approximation is
the local density approximation. However this method also
relies heavily on the variational nature of the ground
state total energy
and this paper is concerned with concepts and
misconcepts related to
this variational property.

Nonvariational nature of the energy functional with variable
occupation numbers has been noticed recently by
Weinert and Davenport \cite{md}. However we will rigorously
prove that the differentiability of the energy functional
with respect to the density depends critically on the choice
of the occupation numbers and that there is no freedom to vary
these  without sacrificing the differentiability of the
energy functional,
i.e. a variational principle does not exist for any arbitrary
choice of occupation numbers within the current framework of the
density functional calculations.  As we show later this means
that Janak's theorem cannot exist within the current framework
of density functional calculations. To our knowledge
the variational nature of the
occupation numbers has never been studied in this fashion.
The significance and relevance
of the above  statements should be judged in the context of the
references made to this problem over
the past two decades \cite{fpn,janak,pz,ap,pj}.

Original Hohenberg-Kohn-Sham theory \cite{hk,ks} (HKS) rests
on the assumption
that the ground state density of the interacting
system is simultaneously a ground state density of the
noninteracting system with suitably chosen external potential.
The ground state for the noninteracting system can always be
obtained
by completely filling the lowest single-particle orbitals.
The occupation numbers in this approach are well defined
(except for
the case when the topmost level is degenerate) and any
other choice would
be in direct violation with the original HKS
assumption. In order
to make any rigorous statements about this particular choice
for the occupation numbers one has to define more
general energy
functional with the domain that includes densities which
are not the ground state densities of any noninteracting
system. Such energy functional
can be defined by using the constrained search approach
proposed by
Levy \cite{lp} .
\section{Constrained search approach.}
Constrained search approach\cite{lp}  defines the
energy functional
in a very simple and  a physically appealing way. The
ground state of any many-electron system can be found as
\begin{equation}
E_0=\min_{\Psi}\langle\Psi|\hat{H}|\Psi\rangle .
\end{equation}
Here the minimization is performed over all allowable
antisymmetric wave functions. To include mixed states
as well as pure states one can extend the minimization
from wave functions to all allowable density matrices:
\begin{equation}
E_0=\min_{\hat{\Gamma}}Tr\{\hat{\Gamma}\hat{H}\} ,
\end{equation}
where
\begin{eqnarray}
\hat{\Gamma}=\sum d_i|\Psi_i\rangle\langle\Psi_i| ,
\\ \nonumber \\
d_i=d_i^{\star}\geq 0 \:,\:\sum d_i=1 \nonumber .
\end{eqnarray}
The minimization can be divided into two steps:
\begin{equation}
E_0=\min_n\{\min_{\hat{\Gamma}\rightarrow n}Tr
\{\hat{\Gamma}\hat{H}\}\} .
\end{equation}
The first minimization is performed over all
allowable density matrices that lead to some
fixed density $n({\bf r})$. The result of this
minimization is the functional of the density only:
\begin{equation}
E_{L}[n]=\min_{\hat{\Gamma}\rightarrow n}Tr
\{\hat{\Gamma}\hat{H}\} .
\end{equation}
According to eq. (4) once we have found this
functional, the ground state energy as well as the
ground state density can be found by minimizing eq. (5)
with respect to the density, i.e.
\begin{equation}
E_0=\min_{n}E_{L}[n] .
\end{equation}
The Hamiltonian for the many-electron system in general
can be written as
\begin{equation}
\hat{H}=\hat{T}+\hat{W}+\hat{V} ,
\end{equation}
where
\begin{equation}
\hat{T}=-\frac{1}{2}\int{\bf \psi}^{\dag}({\bf r})
\nabla^2 {\bf \psi}({\bf r}) d{\bf r}
\end{equation}
is the kinetic energy term,
\begin{equation}
\hat{W}=\frac{1}{2}\int \frac{1}{|{\bf r}-{\bf r}^{\prime}|}
\psi^{\dag}({\bf r})\psi^{\dag}({\bf r}^{\prime})\psi({
\bf r}^{\prime})\psi({\bf r})d {\bf r}
\end{equation}
is the electron-electron interaction term,
\begin{equation}
\hat{V}=\int v({\bf r}) \psi^{\dag}({\bf r})\psi(
{\bf r}) d{\bf r}
\end{equation}
is the external field interaction term.
According to eq.(5) this leads to
\begin{equation}
E_{L}[n]=F_{L}[n]+\int v({\bf r})n({\bf r})d{\bf r} .
\end{equation}
All the complexities of many-electron system are now
hidden in the functional $F_{L}[n]$:
\begin{equation}
F_{L}[n]=\min_{\hat{\Gamma}\rightarrow n}Tr\{\hat{\Gamma}\hat{T}
+\hat{\Gamma}\hat{W}\} .
\end{equation}
It was shown by Lieb \cite{lieb} that $F_{L}[n]$
and therefore
$E_{L}[n]$ are convex functionals.
This means that $E_{L}[n]$ has no extrema above
its absolute minimum
\cite{levy}.

The following important theorem by H.Englisch
and R.Englisch \cite{eng} is
central to our conclusion and deals with the issue
of the differentiability
of $F_{L}[n]$.

{\bf Theorem 1:}
{\it
The functional \[ F_{L}[n]=\min_{\hat{\Gamma}\rightarrow n}
Tr\{\hat{\Gamma}\hat{T}+\hat{\Gamma}\hat{W}\} \]
is differentiable
nowhere else but on the following set of densities:
\[ n({\bf r})=\sum_{k}\lambda_{k} n_{k}({\bf r}) ,\]
with $\sum_{k}\lambda_{k}=1$, $\lambda_{k}\geq 0$,
and $n_{k}({\bf r})=
\int|\psi_{k}({\bf r},{\bf r}_2,...,{\bf r}_N)|^
{2}d{\bf r}_2...d{\bf r}_N $,
 where $\psi_{k}$ are the orthonormal, degenerate
ground states
of the Hamiltonian given by \rm{(7)}. Moreover the
choice for the external potential
is unique (up to some additive constant) for a given density.
}

As a corollary to the above theorem one can prove the
following statement:

{\bf Proposition 1:}
{\it The functional
\begin{equation}
T_{J}[n]=\min_{\sum_i f_i|\phi_i|^2\rightarrow n}
\sum_i f_i\int {\phi}^{\star}_{i}({\bf r})\left
(-\frac{\nabla^2}{2}\right )\phi_i({\bf r}) d{\bf r}
\end{equation}
is differentiable nowhere else but on the following set of densities:
\[ n({\bf r})=\sum_i f_i|\phi_i({\bf r})|^2 , \]
where
\[ \left(-\frac{\nabla^2}{2}+v({\bf r})\right)\phi_i({\bf r})=
\epsilon_{i}\phi_i({\bf r}) ,\]  and $\sum f_i=N$ ,
\begin{equation}
f_i=\left\{ \begin{array}{lll}
1 & \mbox{for $\epsilon_i<\epsilon_F$ }\\
0  & \mbox{for $\epsilon_i>\epsilon_F$ }\\
x &  \mbox{for $\epsilon_i=\epsilon_F$ , where $0\leq x \leq 1$}
\end{array} \right .
\end{equation}
}
The proof of the above statement follows from the fact that
ensemble-N-representable first order density matrix (i.e
derivable from N-particle density matrix $\Gamma$) can
always be written as \cite{lp,col,dft2}
\begin{equation}
\gamma({\bf r},{\bf r}^{\prime})=\sum_i f_i\phi_i^{\star}({\bf
r}^{\prime})\phi_i({\bf r}) ,
\end{equation}
where $ 0\leq f_i\leq 1$ , $\sum_i f_i=N $ and $\phi_i$'s
are orthonormal.
Since the expectation value of the kinetic energy
depends only on the first order density matrix it
follows that \cite{pz}
\[ T_{J}[n]=\min_{\Gamma\rightarrow n}Tr\{\hat{
\Gamma}\hat{T}\} . \]
Once this equivalence has been established, we can now
apply Theorem 1 (with $\hat{W}$ set to zero) to the functional
$ T_{J}[n]$.
In this case $\psi_{k}$'s are the ground states of the following
Hamiltonian
\[ H_{0}=-\sum_{i=1}^{N}\frac{1}{2}\nabla^{2}_{i} +
\sum_{i=1}^{N} v({\bf r}_{i}) \]
and can be represented as single determinants built from N lowest
single particle orbitals $\phi({\bf r})$:
\[ (-\frac{1}{2}\nabla^{2}+ v({\bf r}))\phi_{i}
({\bf r})=\epsilon_{i}\phi_{i}
({\bf r}) \;\;\;\;\;\;\; \epsilon_1\leq\epsilon_2
\leq\epsilon_3\leq ... . \]
The only case where we have degenerate many-particle
ground state is
when there is a degenerate single particle level $\epsilon_k$
($k\leq N$) with degeneracy \cite{eng} $m>N-k+1$.
 This energy level $\epsilon_k$
is the Fermi level ($\epsilon_k=\epsilon_f$). Then
\[n_{k}({\bf r})=\int|\psi_{k}({\bf r},{\bf r}_2,...,
{\bf r}_N)|^{2}d{\bf r}_2...d{\bf
r}_N=\sum_{\epsilon_{i}<\epsilon_{F}}|\phi_{i}({\bf r})|^2
+\sum_{\epsilon_{i_{k}}=\epsilon_{F}}|\phi_{i_{k}}({\bf r})|^2 .\]
The densities where the functional $T_{J}[n]$ is differentiable
are given by:
\begin{eqnarray}
n({\bf r})&=&\sum_{k}\lambda_{k} n_{k}({\bf r})=
\sum_{k}\lambda_{k}\sum_{\epsilon_{i}<\epsilon_{F}}|
\phi_{i}({\bf r})|^2
+\sum_{k}\lambda_{k}\sum_{\epsilon_{i_{k}}=\epsilon_{F}}
|\phi_{i_{k}}({\bf r})|^2 \\ \nonumber
&=&\sum_{\epsilon_{i}<\epsilon_{F}}|\phi_{i}({\bf r})|^2
+\sum_{\epsilon_{i}=\epsilon_{F}}f_{i}|\phi_{i}
({\bf r})|^2 ,\nonumber
\end{eqnarray}
where
\[ f_{i}=\sum_{k}\delta_{i_{k},i}\lambda_{k}\leq 1
\;\;\;\;\;\;\; Q.E.D.\]
\section{Janak's theorem.}
It is obvious that the energy functional (11) is of little help when
it comes to practical applications. Further simplifications are needed.
Following the HKS strategy one usually
partitions \cite{pz,dft2,dft1} the functional $F_{L}[n]$ in the following
fashion:
\begin{equation}
F_{L}[n]=T_{J}[n]+\frac{1}{2}\int \int \frac{n({\bf r})n({\bf
r}^{\prime})}{|{\bf r}-{\bf r}^{\prime}|} d{\bf r}d{\bf r}^{\prime}
+ E_{Lxc}[n] .
\end{equation}
This results in the following energy functional
\begin{equation}
E[n]=T_{J}[n]+\int v({\bf r})n({\bf r})d{\bf r}+\frac{1}
{2}\int \int \frac{n({\bf r})n({\bf r}^{\prime})}{|
{\bf r}-{\bf r}^{\prime}|} d{\bf r}d{\bf r}^{\prime} + E_{Lxc}[n] .
\end{equation}
If the above functional was differentiable for any choice of the
occupation numbers, its subsequent minimization with respect
to density
would lead \cite{pz,dft2,dft1} to the following results:
\begin{equation}
\left(-\frac{\nabla^2}{2}+v_{eff}({\bf r})\right)\phi_i({\bf r})=
\epsilon_{i}\phi_i({\bf r}) ,
\end{equation}
where
\[ v_{eff}({\bf r})=v({\bf r})+\int \frac{n({\bf r}^{\prime})}
{|{\bf r}-{\bf r}^{\prime}|}d{\bf r}^{\prime}+
\frac{\delta E_{Lxc}[n]}{\delta n({\bf r})}\;\;\;,\:\:\;\;
n({\bf r})=\sum_i f_i|\phi_i({\bf r})|^2 ,\]
and
\begin{equation}
\frac{\delta E[n]}{\delta f_{i}}=\epsilon_{i} .
\end{equation}
The last result was first obtained by Janak \cite{janak}
before the advent
of the constrained search formulation and
is known as Janak's theorem.
This theorem is usually regarded as a justification
for occupying
only the single particle orbitals below the Fermi
level \cite{pz,pj}.

Nevertheless, the flaw in the above derivation is
obvious. According to
Proposition 1, once the partition (17)
has been performed we have immediate restrictions on
the occupation numbers. Any choice other than
the conventional one,
makes functionals $T_{J}[n]$ and $E_{Lxc}[n]$
(since $E_{Lxc}[n]$ implicitly
contains the kinetic energy term $T_{J}[n]$)
nondifferentiable with respect to density.
Therefore we cannot vary the energy functional (18)
for some arbitrary choice of the of the occupation numbers $\{f_i\}$.
Hence equation (19) is not valid for the occupation numbers other
than the ones specified in Proposition 1. Obviously, in this
case eq. (20) is not valid since we no longer have the freedom
to choose arbitrary occupation numbers. In particular this means
that there
is no room for the Janak's theorem within the current framework
of density functional calculations.

 The question now arises how to occupy the topmost degenerate
levels (if any). As previously mentioned, the functional $E_{L}[n]$ is
convex and there are no extrema above its absolute minimum.
Therefore any fractional occupations at the Fermi level that
conserve the degeneracy (i.e. the functionals $T_{J}[n]$ and
$E_{Lxc}[n]$ remain differentiable) and leave eq. (17)
intact (i.e. deliver the extremum of the functional $E_{L}[n]$ )
should give the same total energy \cite{eng}. This of course  may
not be true in practice because of our approximations to the
exchange-correlation energy.
\section{Conclusion.}
 The functionals $T_{J}[n]$ and $E_{Lxc}[n]$ appear as
a direct consequence of the attempt to map the interacting system
into the noninteracting one. We have demonstrated that these functionals
are differentiable only on a certain domain of densities. It is only
there that we are
allowed to vary the energy functional. It is shown that in order
for density to belong to this domain there is a certain restriction
on the occupation numbers, i.e. single-particle orbitals
below the Fermi level are completely occupied (with possible fractional
occupations at the Fermi level) and the orbitals
above the Fermi level are empty. Any other choice of
the occupation numbers
will make the functionals $T_{J}[n]$ and $E_{Lxc}[n]$ lose their
differentiability with respect to density -- a property which is
imperative for the variational principle. Note that
even though the approximate forms for the exchange-correlation energy
currently being used might appear differentiable with respect to
density regardless of the choice of the occupation numbers, one can
hardly claim that these
approximations will hold for an arbitrary
set of the occupation numbers.

 The main conclusion is that the occupation numbers cannnot
 be considered as variational parameters. It means
 that there is no apriory justification
for using the conventional set of the occupation numbers in the current
density
functional calculations. The only reason we resort to this
set of the occupation numbers is because any other choice
would make the variational
principle inapplicable.
\section{Acknowledgments}
This research was supported in part by DOE Grant\# DE-FG02-87ER45318 and
the University of Connecticut Research Foundation.

\end{document}